# A Continuum Theory of Elastic-Ferromagnetic Conductors

Jiashi Yang
(jyang1@unl.edu)
Department of Mechanical and Materials Engineering
University of Nebraska-Lincoln
Lincoln, NE 68588-0526, USA

**Abstract**

In this paper, a phenomenological theory of saturated ferromagnetoelastic conductors is established using a multi-continuum model and the classical laws of mechanics, thermodynamics and electromagnetics. The theory is nonlinear and is valid for large deformations and strong electromagnetic fields. The constitutive relations in the theory satisfy the saturation condition of the magnetization vector. The theory is with full electromagnetic couplings as governed by Maxwell's equations. It can describe the interactions of elastic, electromagnetic and spin waves. The theory can be reduced to various quasistatic theories with appropriate approximations of the electromagnetic fields. It is for anisotropic materials in general.

**Keywords**: ferromagnetic; elastic; conductor; saturation; photon-phonon-magnon interaction

## 1. Multi-Continuum Model

Various continuum theories for saturated ferromagnetic solids can be found in the literature for both rigid [1-5] and elastic [6-13] materials. Specifically, for elastic and saturated ferromagnets, [6-13] are all or insulators. In this paper we construct a continuum theory for elastic and saturated ferromagnetic conductors using the multi-continuum model which has been used in various forms in [4-8] and [14-18]. Four charged or magnetized continua are needed for the elastic and saturated ferromagnets in the present paper.

In the reference state without any deformation and fields at $t_0$, the reference position of a material point of the lattice continuum [14] in Fig. 1 is denoted by **X**. The mass density and the positive charge density of the lattice continuum at the reference state are $\rho^0(\mathbf{X})$ and $_0\sigma^l(\mathbf{X})$, respectively. At an arbitrary time $t$, the lattice continuum occupies a spatial region $v$ with a boundary surface $s$ whose outward unit normal is **n**. The present position, mass density and charge density of the material point associated with **X** are denoted by **y**, $\rho$ and $\sigma^l$. The motion of the lattice continuum is described by $\mathbf{y}=\mathbf{y}(\mathbf{X},t)$. The velocity field of the lattice continuum is written as $\mathbf{v}(\mathbf{y},t)$. We have the following equations from the conservation of mass [19] and charge for the lattice continuum, respectively:

$$\frac{d\rho}{dt} + \rho\frac{\partial v_k}{\partial y_k} = \dot{\rho} + \rho v_{k,k} = \dot{\rho} + \rho\nabla\cdot\mathbf{v} = 0,$$

$$\frac{d}{dt} = \frac{\partial}{\partial t}\bigg|_{\mathbf{y}} + \mathbf{v}\cdot\nabla,$$

(1.1)

$$\frac{d\sigma^l}{dt} + \sigma^l v_{k,k} = 0.$$

(1.2)

Figure 1 also shows the present state of the lattice continuum with various external electromechanical loads and its internal interactions with the other three continua in Figs. 2, 3 and 5. In the figure, couple or moment vectors are shown by double arrows to distinguish them from force vectors in single arrows. The mechanical forces acting on the lattice continuum are the usual surface traction **t** and body force **f**. The present lattice charge density $\sigma^l(\mathbf{y})$ experiences a force under the Maxwellian electric field **E** and magnetic induction **B**. The interaction between the lattice continuum and the bound charge continuum in Fig. 2 is through an internal electric field $\mathbf{E}^b$. The interactions between the lattice continuum and the spin continuum in Fig. 3 include an internal force $\mathbf{f}^L$ and an internal couple $\mathbf{c}^L$ due to an internal magnetic induction $\mathbf{B}^L$. The interaction between the lattice continuum and the free charge continuum in Fig. 5 is by an effective electric field $\mathbf{E}^e$.

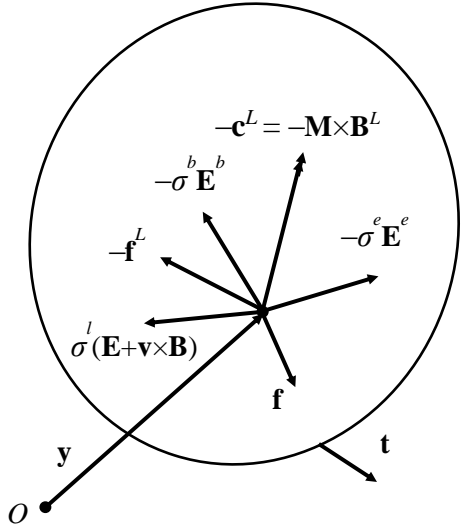

Fig. 1. Current configuration of the lattice continuum.

The bound charge continuum [14] in Fig. 2 is introduced to describe electric polarization. It is massless and has a negative charge density $_0\sigma^b(\mathbf{X})$ at the reference state. The residual charge density $\sigma^r(\mathbf{X})$ from both the lattice and bond charge continua at the reference state is

$$_0\sigma^l(\mathbf{X}) + {_0\sigma^b}(\mathbf{X}) = \sigma^r(\mathbf{X}). \tag{1.3}$$

The bound charge continuum can move slightly from the lattice continuum through a small displacement field denoted by $\mathbf{\eta}(\mathbf{y},t)$. We assume that $\mathbf{\eta}$ satisfies [14]

$$\eta_{k,k} = 0, \tag{1.4}$$

which is sufficient to describe polarization and, at the same time, preserves the form of Gauss's law of the electric field as to be shown later. In the current state, the bound charge $\sigma^b(\mathbf{y}+\mathbf{\eta})$ experiences a force under the Maxwellian electric field $\mathbf{E}$ and magnetic induction $\mathbf{B}$ as shown in Fig. 2. The residue charge at the present state is given by

$$\sigma^l(\mathbf{y},t) + \sigma^b(\mathbf{y}+\mathbf{\eta},t) = \sigma^r(\mathbf{y},t). \tag{1.5}$$

The continuity equations (conservation of charge) for the bound charge and the residual charge are, respectively,

$$\frac{d\sigma^b}{dt} + \sigma^b(v_k + \dot{\eta}_k)_{,k} = \dot{\sigma}^b + \sigma^b v_{k,k} = 0, \tag{1.6}$$

$$\frac{d\sigma^r}{dt} + \sigma^r v_{k,k} = 0. \tag{1.7}$$

(1.1), (1.2) and (1.6) imply that

$$\frac{\dot{\sigma}^l}{\sigma^l} = \frac{\dot{\sigma}^b}{\sigma^b} = \frac{\dot{\rho}}{\rho} = -v_{k,k}. \tag{1.8}$$

The electric polarization per unit volume can be expressed through $\mathbf{\eta}$ by [14]

$$\mathbf{P} = \sigma^l(\mathbf{y})(-\mathbf{\eta}). \tag{1.9}$$

For later use we introduce the polarization per unit mass by

$$\boldsymbol{\pi} = \frac{\mathbf{P}}{\rho}, \tag{1.10}$$

which is more convenient for the polarization of deformable bodies. We have

$$\sigma^b \dot{\boldsymbol{\eta}} = \frac{d}{dt}\left(\sigma^b \boldsymbol{\eta}\right) - \dot{\sigma}^b \boldsymbol{\eta} = \dot{\mathbf{P}} - \frac{\mathbf{P}}{\sigma^b}\dot{\sigma}^b$$

$$= \dot{\mathbf{P}} - \mathbf{P}\frac{\dot{\rho}}{\rho} = \rho\dot{\boldsymbol{\pi}} + \dot{\rho}\boldsymbol{\pi} - \dot{\rho}\boldsymbol{\pi} = \rho\dot{\boldsymbol{\pi}}, \tag{1.11}$$

where (1.8) has been used. (1.11) which will be used in several places in later derivations.

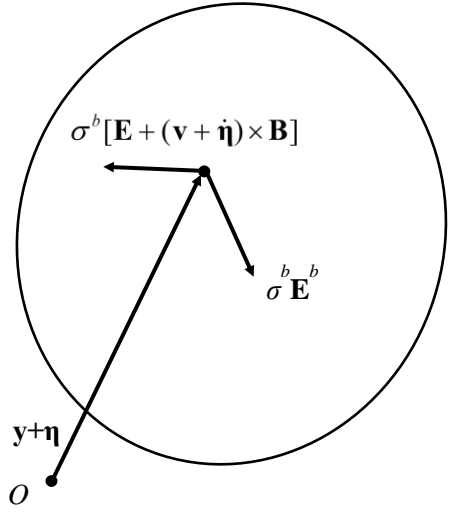

Fig. 2. Current configuration of the bound charge continuum.

The spin continuum [6] is shown in Fig. 3 along with the forces and couples acting on it. The spin continuum is massless. It carries distributed circulating currents which form a magnetic moment field $\mathbf{M}'(\mathbf{y},t)$ where the prime indicates that it is a field in the co-moving or instantaneous rest frame associated with the material particle of the lattice at $\mathbf{y}$ [8]. The spin continuum is assumed to be fixed to the lattice continuum and cannot move with respect to the lattice continuum. However, $\mathbf{M}'$ can rotate with respect to the lattice through some angular motion. The force $\mathbf{f}^M$ and couple $\mathbf{c}^M$ on the spin continuum due to the Maxwellian magnetic induction $\mathbf{B}$ and the internal couple $\mathbf{c}^L$ have the following expressions [8]:

$$\mathbf{f}^M = \mathbf{M}' \cdot (\mathbf{B}\nabla), \quad \mathbf{c}^M = \mathbf{M}' \times \mathbf{B},$$
$$\mathbf{c}^L = \mathbf{M}' \times \mathbf{B}^L. \tag{1.12}$$

On the surface of the spin continuum, it is assumed that there exists a continuous distribution of couples in the form of $\mathbf{M}' \times \mathbf{F}$ per unit area due to an exchange field $\mathbf{F}$ which is from a quantum mechanical origin. We note that $\mathbf{F}$ and $\mathbf{B}^L$ act on $\mathbf{M}'$ through cross products. Therefore, without loss of generality, we assume that [6,8]

$$\mathbf{F} \cdot \mathbf{M}' = 0,$$
$$\mathbf{B}^L \cdot \mathbf{M}' = 0. \tag{1.13}$$

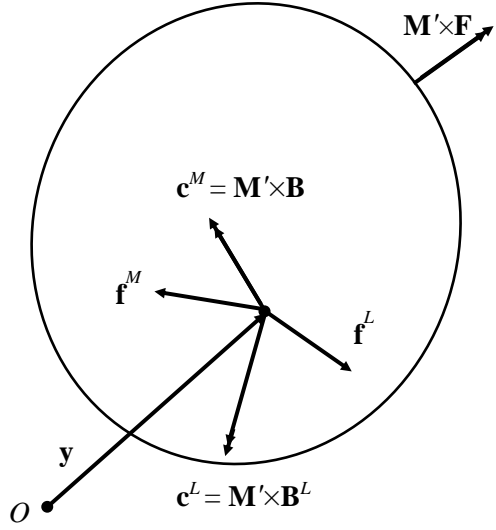

Fig. 3. Current configuration of the spin continuum.

For later use we introduce the magnetization per unit mass by

$$\boldsymbol{\mu}' = \frac{\mathbf{M}'}{\rho}. \tag{1.14}$$

Then the magnetic saturation condition can be written as

$$\boldsymbol{\mu}' \cdot \boldsymbol{\mu}' = \mu_s'^2, \tag{1.15}$$

where $\mu_s'$ is the saturation magnetization. Differentiating (1.15) with respect to the time $t$ and/or the reference coordinates $\mathbf{X}$, we obtain

$$\mu_k' \frac{d\mu_k'}{dt} = 0, \quad \mu_k' \mu_{k,L}' = 0, \quad \mu_{k,L}' \frac{d\mu_k'}{dt} + \mu_k' \frac{d\mu_{k,L}'}{dt} = 0. \tag{1.16}$$

While the magnitude of $\boldsymbol{\mu}'$ is a constant as restricted by the saturation condition, the direction of $\boldsymbol{\mu}'$ can change. The rotation or angular displacement of the direction of $\boldsymbol{\mu}'$ can be described by $\delta\theta = |\delta\boldsymbol{\mu}'|/|\boldsymbol{\mu}'|$ as shown in Fig. 4.

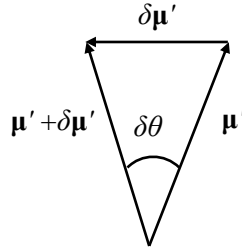

Fig. 4. Change of direction of $\boldsymbol{\mu}'$.

For a complete description of the change of direction of $\boldsymbol{\mu}'$ we introduce an angular displacement vector $\boldsymbol{\delta\theta}$ by

$$\boldsymbol{\delta\theta} = \delta\theta \frac{\boldsymbol{\mu}'}{|\boldsymbol{\mu}'|} \times \frac{\delta\boldsymbol{\mu}'}{|\delta\boldsymbol{\mu}'|} = \frac{|\delta\boldsymbol{\mu}'|}{|\boldsymbol{\mu}'|} \frac{\boldsymbol{\mu}'}{|\boldsymbol{\mu}'|} \times \frac{\delta\boldsymbol{\mu}'}{|\delta\boldsymbol{\mu}'|} = \frac{1}{\mu_s'^2} \boldsymbol{\mu}' \times (\delta\boldsymbol{\mu}'). \tag{1.17}$$

Then it can be shown that [6,8]

$$\boldsymbol{\delta\theta}\times\boldsymbol{\mu}' = \frac{1}{\mu_s'^2}(\boldsymbol{\mu}'\times\delta\boldsymbol{\mu}')\times\boldsymbol{\mu}' = \frac{1}{\mu_s'^2}\left[(\boldsymbol{\mu}'\cdot\boldsymbol{\mu}')\delta\boldsymbol{\mu}' - (\boldsymbol{\mu}'\cdot\delta\boldsymbol{\mu}')\boldsymbol{\mu}'\right] = \delta\boldsymbol{\mu}'. \quad (1.18)$$

Corresponding to $\boldsymbol{\delta\theta}$, we also introduce an angular velocity vector for the change of direction of $\boldsymbol{\mu}'$ by

$$\boldsymbol{\omega} = \lim_{\delta t \to 0}\frac{\boldsymbol{\delta\theta}}{\delta t} = \frac{1}{\mu_s'^2}\boldsymbol{\mu}'\times\frac{d\boldsymbol{\mu}'}{dt}. \quad (1.19)$$

The power of a magnetic couple in the form of $\boldsymbol{\Gamma} = \rho\boldsymbol{\mu}'\times\mathbf{B}$ on the magnetic moment $\rho\boldsymbol{\mu}'$ can be calculated as [6,8]

$$w^M = \boldsymbol{\Gamma}\cdot\boldsymbol{\omega} = (\rho\boldsymbol{\mu}'\times\mathbf{B})\cdot\left(\frac{1}{\mu_s'^2}\boldsymbol{\mu}'\times\frac{d\boldsymbol{\mu}'}{dt}\right) = \frac{\rho}{\mu_s'^2}\left[(\boldsymbol{\mu}'\cdot\boldsymbol{\mu}')\left(\mathbf{B}\cdot\frac{d\boldsymbol{\mu}'}{dt}\right) - \left(\boldsymbol{\mu}'\cdot\frac{d\boldsymbol{\mu}'}{dt}\right)(\mathbf{B}\cdot\boldsymbol{\mu}')\right] = \rho\mathbf{B}\cdot\frac{d\boldsymbol{\mu}'}{dt}. \quad (1.20)$$

The angular momentum of $\mathbf{M}'=\rho\boldsymbol{\mu}'$ is given by

$$\mathbf{L}' = \frac{1}{\gamma}\mathbf{M}', \quad (1.21)$$

where $\gamma$ is the gyromagnetic ratio, a negative number.

The free charge continuum [18] in Fig. 5 is also assumed to be massless. Its charge density and velocity field are denoted by $\sigma^e(\mathbf{y},t)$ and $\mathbf{v}^e(\mathbf{y},t)$. The force on the free charge continuum due to the Maxwellian electric field $\mathbf{E}$ and magnetic induction $\mathbf{B}$ is as shown in the figure, along with its interaction with the lattice continuum. The continuity equation for the free charge continuum is

$$\frac{\partial\sigma^e}{\partial t} + (\sigma^e v_k^e)_{,k} = 0. \quad (1.22)$$

The total charge density and total current density of the combined continuum are given by [18]

$$\sigma = \sigma^r + \sigma^e, \quad \mathbf{J} = \sigma^r\mathbf{v} + \sigma^e\mathbf{v}^e, \quad (1.23)$$

which satisfy

$$\frac{\partial\sigma}{\partial t} + J_{k,k} = 0. \quad (1.24)$$

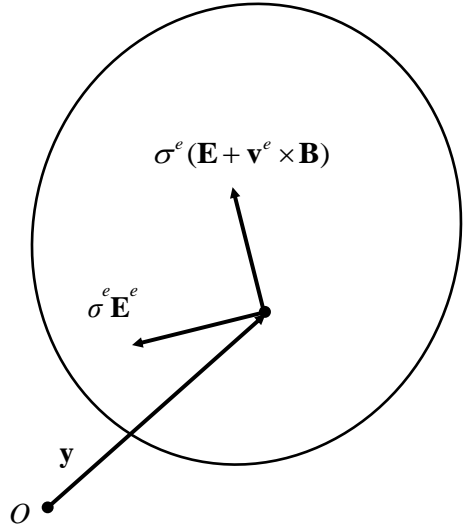

Fig. 5. Current configuration of the free charge fluid.

With the lattice charge, bound charge and free charge, we can write the integral form of Gauss' law for the electric field as [16]

$$\oint_s \varepsilon_0 \mathbf{n} \cdot \mathbf{E} ds = \int_v [\sigma^l(\mathbf{y}) + \sigma^b(\mathbf{y}) + \sigma^e(\mathbf{y})] dv, \qquad (1.25)$$

whose differential form is

$$\varepsilon_0 E_{k,k} = -P_{i,i} + \sigma, \qquad (1.26)$$

or

$$(\varepsilon_0 E_i + P_i)_{,i} = D_{i,i} = \sigma,$$
$$D_i = \varepsilon_0 E_i + P_i, \qquad (1.27)$$

where $\mathbf{D}$ the electric displacement vector.

## 2. Electromagnetic Body Force, Couple and Power

Consider a unit volume of the lattice continuum, spin continuum and free charge fluid together at $\mathbf{y}$. The corresponding bound charge continuum at $\mathbf{y}+\boldsymbol{\eta}$ also has a unit volume. From Figs. 1, 2, 3 and 5, the electromagnetic body force on a unit volume of the combined continuum of four is

$$\mathbf{F}^{EM} = \sigma^l(\mathbf{y})[\mathbf{E}(\mathbf{y}) + \mathbf{v}(\mathbf{y}) \times \mathbf{B}(\mathbf{y})] + \sigma^b(\mathbf{y}+\boldsymbol{\eta})[\mathbf{E}(\mathbf{y}+\boldsymbol{\eta}) + (\mathbf{v}+\dot{\boldsymbol{\eta}}) \times \mathbf{B}(\mathbf{y}+\boldsymbol{\eta})]$$
$$+ \mathbf{M}' \cdot (\mathbf{B}\nabla) + \sigma^e(\mathbf{y})[\mathbf{E}(\mathbf{y}) + \mathbf{v}^e(\mathbf{y}) \times \mathbf{B}(\mathbf{y})]. \qquad (2.1)$$

Using the following approximations [8],

$$E_j(\mathbf{y}+\boldsymbol{\eta}) \cong E_j(\mathbf{y}) + \eta_i E_{j,i}(\mathbf{y}),$$
$$B_j(\mathbf{y}+\boldsymbol{\eta}) \cong B_j(\mathbf{y}) + \eta_i B_{j,i}(\mathbf{y}), \qquad (2.2)$$

through some lengthy algebra, we can write $\mathbf{F}^{EM}$ as

$$\mathbf{F}^{EM} = \mathbf{P} \cdot \nabla \mathbf{E} + \mathbf{M}' \cdot (\mathbf{B}\nabla) + \mathbf{v} \times (\mathbf{P} \cdot \nabla \mathbf{B}) + \rho \dot{\boldsymbol{\pi}} \times \mathbf{B} + \sigma \mathbf{E} + \mathbf{J} \times \mathbf{B}, \qquad (2.3)$$

where a term of the product of the small $\boldsymbol{\eta}$, its material time derivative and the gradient of $\mathbf{B}$ has been omitted as an approximation [8]. It can be shown that [8]

$$F_j^{EM} = T_{ij,i}^{EM} - \frac{\partial G_j}{\partial t}, \qquad (2.4)$$

where

$$\mathbf{G} = \varepsilon_0 \mathbf{E} \times \mathbf{B}, \quad G_j = \varepsilon_0 \varepsilon_{jkl} E_k B_l \qquad (2.5)$$

is the electromagnetic momentum density and

$$T_{ij}^{EM} = P_i E_j' - B_i M_j' + \varepsilon_0 E_i E_j + \frac{1}{\mu_0} B_i B_j - \frac{1}{2}\left(\varepsilon_0 E_k E_k + \frac{1}{\mu_0} B_k B_k - 2M_k' B_k\right)\delta_{ij} \qquad (2.6)$$

is the electromagnetic stress tensor.

The electromagnetic couple on a unit volume of the combined continuum about $\mathbf{y}=0$ is given by

$$\mathbf{y} \times \left[\sigma^l(\mathbf{y})\mathbf{E}(\mathbf{y}) + \sigma^l(\mathbf{y})\mathbf{v}(\mathbf{y}) \times \mathbf{B}(\mathbf{y}) + \mathbf{M}' \cdot (\mathbf{B}\nabla) + \sigma^e(\mathbf{y})\mathbf{E}(\mathbf{y}) + \sigma^e(\mathbf{y})\mathbf{v}^e(\mathbf{y}) \times \mathbf{B}(\mathbf{y})\right]$$
$$+ (\mathbf{y}+\boldsymbol{\eta}) \times \left[\sigma^b(\mathbf{y}+\boldsymbol{\eta})\mathbf{E}(\mathbf{y}+\boldsymbol{\eta}) + \sigma^b(\mathbf{y}+\boldsymbol{\eta})(\mathbf{v}+\dot{\boldsymbol{\eta}}) \times \mathbf{B}(\mathbf{y}+\boldsymbol{\eta})\right] + \mathbf{M}' \times \mathbf{B} \qquad (2.7)$$
$$= \mathbf{y} \times \mathbf{F}^{EM} + \mathbf{C}^{EM},$$

where we have introduced the electromagnetic body couple $\mathbf{C}^{EM}$ per unit volume by

$$\mathbf{C}^{EM} = \mathbf{P} \times \mathbf{E}' + \mathbf{M}' \times \mathbf{B}, \qquad (2.8)$$
$$\mathbf{E}' = \mathbf{E} + \mathbf{v} \times \mathbf{B}, \quad \mathbf{M}' \cong \mathbf{M} + \mathbf{v} \times \mathbf{P}. \qquad (2.9)$$

The electromagnetic power per unit volume of the combined continuum of four is

$$W^{EM} = \sigma^l(\mathbf{y})[\mathbf{E}(\mathbf{y}) + \mathbf{v}(\mathbf{y}) \times \mathbf{B}(\mathbf{y})] \cdot \mathbf{v} + \sigma^e(\mathbf{y})[\mathbf{E}(\mathbf{y}) + \mathbf{v}^e(\mathbf{y}) \times \mathbf{B}(\mathbf{y})] \cdot \mathbf{v}^e$$
$$+ \sigma^b(\mathbf{y}+\boldsymbol{\eta})[\mathbf{E}(\mathbf{y}+\boldsymbol{\eta}) + (\mathbf{v}+\dot{\boldsymbol{\eta}}) \times \mathbf{B}(\mathbf{y}+\boldsymbol{\eta})] \cdot (\mathbf{v}+\dot{\boldsymbol{\eta}}) - \mathbf{M}' \cdot \frac{\partial \mathbf{B}}{\partial t} \quad (2.10)$$
$$= (\mathbf{P} \cdot \nabla \mathbf{E}) \cdot \mathbf{v} + \rho \mathbf{E} \cdot \dot{\boldsymbol{\pi}} - \mathbf{M}' \cdot \frac{\partial \mathbf{B}}{\partial t} + \mathbf{E} \cdot \mathbf{J}$$
$$= \mathbf{F}^{EM} \cdot \mathbf{v} + \rho \mathbf{E}' \cdot \dot{\boldsymbol{\pi}} - \mathbf{M}' \cdot \dot{\mathbf{B}} + \mathbf{J}' \cdot \mathbf{E}',$$

where

$$\mathbf{J}' = \mathbf{J} - \sigma \mathbf{v} = \sigma^e (\mathbf{v}^e - \mathbf{v}). \quad (2.11)$$

## 3. Global Balance Laws

The global or integral forms of Maxwell's equations are the same as those in [5]. The linear momentum equation and the angular momentum equation about $\mathbf{y}=0$ for the spin continuum alone are

$$\int_v (\mathbf{f}^M + \mathbf{f}^L) dv = 0, \quad (3.1)$$

$$\frac{d}{dt}\int_v \rho \frac{\boldsymbol{\mu}'}{\gamma} dv = \int_v \mathbf{y} \times (\mathbf{f}^M + \mathbf{f}^L) dv + \int_s \rho \boldsymbol{\mu}' \times \mathbf{F} ds + \int_v \rho \boldsymbol{\mu}' \times (\mathbf{B} + \mathbf{B}^L) dv. \quad (3.2)$$

For the combined continuum, let $\varepsilon$ be the internal energy density per unit mass, $r$ the body heat source per unit mass, $\mathbf{q}$ the heat flux vector, $\theta$ the absolute temperature and $\eta$ the entropy density per unit mass. Then the conservation of mass, the linear and angular momentum equations, the energy equation and the second law of thermodynamics can be written as

$$\frac{d}{dt}\int_v \rho dv = 0, \quad (3.3)$$

$$\frac{d}{dt}\int_v \rho \mathbf{v} dv = \int_s \mathbf{t} ds + \int_v (\rho \mathbf{f} + \mathbf{F}^{EM}) dv, \quad (3.4)$$

$$\frac{d}{dt}\int_v \left(\mathbf{y} \times \rho \mathbf{v} + \rho \frac{\boldsymbol{\mu}'}{\gamma}\right) dv = \int_s \mathbf{y} \times \mathbf{t} ds + \int_s \rho \boldsymbol{\mu}' \times \mathbf{F} ds + \int_v \left[\mathbf{y} \times (\rho \mathbf{f} + \mathbf{F}^{EM}) + \mathbf{C}^{EM}\right] dv, \quad (3.5)$$

$$\frac{d}{dt}\int_v \rho\left(\frac{1}{2}\mathbf{v}\cdot\mathbf{v} + \varepsilon\right) dv = \int_s \left(\mathbf{t}\cdot\mathbf{v} - \mathbf{n}\cdot\mathbf{q} + \mathbf{F}\cdot\rho\frac{d\boldsymbol{\mu}'}{dt}\right) ds + \int_v (\rho\mathbf{f}\cdot\mathbf{v} + \rho r) dv + \int_v W^{EM} dv, \quad (3.6)$$

$$\frac{d}{dt}\int_v \rho \eta dv \geq \int_v \frac{\rho r}{\theta} dv - \int_s \frac{\mathbf{q}\cdot\mathbf{n}}{\theta} ds. \quad (3.7)$$

## 4. Local Balance Laws

In this section we convert the global balance laws in the previous section to local or differential forms. The differential forms of Maxwell's equations are the same as those in [5]. Since the spin continuum is massless, its linear momentum equation in (3.1) simply leads to $\mathbf{f}^M + \mathbf{f}^L = 0$. For the angular momentum equation of the spin continuum alone in (3.2), we express the vector $\mathbf{F}$ by a tensor $\mathbf{A}$ through [6]

$$\mathbf{F} = -\mathbf{n}\cdot\mathbf{A}. \quad (4.1)$$

Accordingly, $(1.13)_1$ on $\mathbf{F}$ becomes the following condition on $\mathbf{A}$:

$$\mathbf{A}\cdot\boldsymbol{\mu}' = 0. \quad (4.2)$$

Then, with the use of the divergence theorem, (3.2) can be converted into differential form as:

$$\frac{1}{\gamma}\rho\frac{d\mu'_i}{dt} = \varepsilon_{ijk}\rho\mu'_j\left(-A_{lk,l} - \frac{A_{lk}}{\rho}\rho_{,l} + B_k + B_k^L\right) - \varepsilon_{ijk}\rho A_{lk}\mu'_{j,l}. \quad (4.3)$$

Taking a dot produce of both sides of (4.3) by $\boldsymbol{\mu}'$, we obtain

$$\frac{1}{\gamma}\mu'_i \frac{d\mu'_i}{dt} = -\mu'_i \varepsilon_{ijk} A_{lk} \mu'_{j,l}. \tag{4.4}$$

The left-hand side of (4.4) vanishes according to (1.16)$_1$. Then, what is left from (4.4) is

$$-\mu'_i \varepsilon_{ijk} A_{lk} \mu'_{j,l} = 0. \tag{4.5}$$

To ensure the satisfaction of (4.5), we require that **A** satisfies [6]:

$$A_{lk}\mu'_{j,l} = A_{lj}\mu'_{k,l}. \tag{4.6}$$

Then (4.3) reduces to

$$\frac{1}{\gamma}\frac{d\mu'_i}{dt} = \varepsilon_{ijk}\mu'_j \left( -A_{lk,l} - \frac{A_{lk}}{\rho}\rho_{,l} + B_k + B_k^L \right). \tag{4.7}$$

Next we dot both sides of (4.7) by the following vector in the direction of **ω**:

$$\varepsilon_{imn}\mu'_m \frac{d\mu'_n}{dt}. \tag{4.8}$$

This yields a relationship which will be used later:

$$\frac{d\mu'_k}{dt}\left( -A_{lk,l} - \frac{A_{lk}}{\rho}\rho_{,l} + B_k + B_k^L \right) = 0. \tag{4.9}$$

The conservation of mass in (3.3) leads to

$$\frac{d\rho}{dt} + \rho \nabla \cdot \mathbf{v} = 0. \tag{4.10}$$

The differential form of the linear momentum equation of the combined continuum in (3.4) is [8]

$$\rho \frac{d\mathbf{v}}{dt} = \nabla \cdot \boldsymbol{\tau} + \rho \mathbf{f} + \mathbf{F}^{EM}. \tag{4.11}$$

The angular momentum equation of the combined continuum in (3.5) leads to

$$\frac{d}{dt}\int_v \rho \frac{\boldsymbol{\mu}'}{\gamma} dv = \int_s \rho \boldsymbol{\mu}' \times \mathbf{F} ds + \int_v \mathbf{y} \times \left( \nabla \cdot \boldsymbol{\tau} + \rho \mathbf{f} + \mathbf{F}^{EM} - \rho \dot{\mathbf{v}} \right) dv + \int_v \left( \mathbf{e}_i \varepsilon_{ijk} \tau_{jk} + \mathbf{C}^{EM} \right) dv, \tag{4.12}$$

whose differential form is

$$\frac{1}{\gamma}\rho\frac{d\mu'_i}{dt} = \varepsilon_{ijk}\tau_{jk} + C_i^{EM} + \varepsilon_{ijk}\rho\mu'_j \left( -A_{lk,l} - \frac{A_{lk}}{\rho}\rho_{,l} \right). \tag{4.13}$$

From (4.7) and (4.13), we obtain

$$\varepsilon_{ijk}\tau_{jk} + C_i^{EM} - \varepsilon_{ijk}\rho\mu'_j \left( B_k + B_k^L \right) = 0, \tag{4.14}$$

or, after the use of (2.8),

$$\varepsilon_{ijk}\tau_{jk} + \varepsilon_{ijk}P_j E'_k - \varepsilon_{ijk}\rho\mu'_j B_k^L = 0. \tag{4.15}$$

The differential forms of (3.6) and (3.7) can be obtained using the divergence theorem in tensor calculus as:

$$\rho\frac{d\varepsilon}{dt} = \tau_{ij}v_{j,i} - A_{ij,i}\rho\left(\frac{d\mu'_j}{dt}\right) - A_{ij}\rho_{,i}\left(\frac{d\mu'_j}{dt}\right) - A_{ij}\rho\left(\frac{d\mu'_j}{dt}\right)_{,i} - \rho\mu'_i\frac{dB_i}{dt} + \rho E'_i\frac{d\pi_i}{dt} + J'_i E'_i + \rho r - q_{i,i}, \tag{4.16}$$

$$\rho\frac{d\eta}{dt} \geq \frac{\rho r}{\theta} - \left(\frac{q_i}{\theta}\right)_{,i}. \tag{4.17}$$

## 5. Constitutive Relations

The energy equation in (4.16) assumes the following form with the use of (4.9):

$$\rho\frac{d\varepsilon}{dt} = \tau_{ij}v_{j,i} - A_{ij}\rho\left(\frac{d\mu'_j}{dt}\right)_{,i} - (B_k + B_k^L)\rho\frac{d\mu'_k}{dt} - \rho\mu'_i\frac{dB_i}{dt} + \rho E'_i\frac{d\pi_i}{dt} + J'_i E'_i + \rho r - q_{i,i}. \tag{5.1}$$

To obtain constitutive relations with the electric field and temperature as independent constitutive variables, we introduce the following Legendre transform:

$$F = \varepsilon - E'_i \pi_i + B_i \mu'_i - \theta \eta. \tag{5.2}$$

Then the energy equation in (5.1) becomes

$$\rho\left(\frac{dF}{dt} + \frac{d\theta}{dt}\eta + \theta\frac{d\eta}{dt}\right) = \tau_{ij} v_{j,i} - A_{ij}\rho\left(\frac{d\mu'_j}{dt}\right)_{,i} - B_k^L \rho \frac{d\mu'_k}{dt} - \rho\pi_i \frac{dE'_i}{dt} + J'_i E'_i + \rho r - q_{i,i}. \tag{5.3}$$

Eliminating $r$ from the entropy inequality in (4.17) and the energy equation in (5.3), we obtain the Clausius–Duhem inequality as

$$-\rho\left(\frac{dF}{dt} + \frac{d\theta}{dt}\eta\right) + \tau_{ij} v_{j,i} - A_{ij}\rho\left(\frac{d\mu'_j}{dt}\right)_{,i} - B_k^L \rho \frac{d\mu'_k}{dt} - P_i \frac{dE'_i}{dt} - \frac{q_i}{\theta}\theta_{,i} + J'_i E'_i \geq 0. \tag{5.4}$$

For constitutive relations including both reversible and nonreversible behaviors such as heat and electrical conductions, we break the stress tensor $\boldsymbol{\tau}$, polarization vector $\mathbf{P}$ and local magnetic induction vector $\mathbf{B}^L$ into reversible and dissipative parts as

$$\boldsymbol{\tau} = \boldsymbol{\tau}^R + \boldsymbol{\tau}^D, \quad \mathbf{P} = \mathbf{P}^R + \mathbf{P}^D, \quad \mathbf{B}^L = {}^R\mathbf{B}^L + {}^D\mathbf{B}^L. \tag{5.5}$$

The reversible parts of $\boldsymbol{\tau}$, $\mathbf{P}$ and $\mathbf{B}^L$ are determined from the free energy $F$ through

$$\rho \frac{dF}{dt} = \tau_{ij}^R v_{j,i} - A_{ij}\rho\left(\frac{d\mu'_j}{dt}\right)_{,i} - {}^R B_k^L \rho \frac{d\mu'_k}{dt} - P_i^R \frac{dE'_i}{dt} - \rho\eta \frac{d\theta}{dt}. \tag{5.6}$$

Then what remains from the energy equation in (5.3) and the Clausius–Duhem inequality in (5.4) are the following:

$$\rho\theta \frac{d\eta}{dt} = \tau_{ij}^D v_{j,i} - {}^D B_k^L \rho \frac{d\mu'_k}{dt} - P_i^D \frac{dE'_i}{dt} + J'_i E'_i + \rho r - q_{i,i}, \tag{5.7}$$

$$\tau_{ij}^D v_{j,i} - {}^D B_k^L \rho \frac{d\mu'_k}{dt} - P_i^D \frac{dE'_i}{dt} - \frac{q_i}{\theta}\theta_{,i} + J'_i E' \geq 0. \tag{5.8}$$

(5.7) is the heat or dissipation equation. With the following kinematic relations [19]:

$$v_{j,i} = X_{M,i}\frac{d}{dt}(y_{j,M}), \quad \left(\frac{d\mu'_j}{dt}\right)_{,i} = X_{M,i}\frac{d}{dt}(\mu'_{j,M}). \tag{5.9}$$

we can write (5.6) as

$$\rho\frac{dF}{dt} = \tau_{ij}^R X_{M,i}\frac{d}{dt}(y_{j,M}) - \rho\,{}^R B_j^L \frac{d\mu'_j}{dt} - \rho A_{ij} X_{M,i}\frac{d}{dt}(\mu'_{j,M}) - P_i^R \frac{dE'_i}{dt} - \rho\eta \frac{d\theta}{dt}. \tag{5.10}$$

Motivated by the right-hand side of (5.10), we consider the following form of $F$:

$$F = F(y_{j,M}; \mu'_i; \mu'_{j,M}; E'_i; \theta). \tag{5.11}$$

Then

$$\frac{dF}{dt} = \frac{\partial F}{\partial(y_{j,M})}\frac{d}{dt}(y_{j,M}) + \frac{\partial F}{\partial \mu'_i}\frac{d\mu'_i}{dt} + \frac{\partial F}{\partial(\mu'_{j,M})}\frac{d}{dt}(\mu'_{j,M}) + \frac{\partial F}{\partial E'_i}\frac{dE'_i}{dt} + \frac{\partial F}{\partial \theta}\frac{d\theta}{dt}. \tag{5.12}$$

Substituting (5.12) into (5.10) and using Lagrange multipliers $\lambda$ and $L_M$ to introduce the constrains in $(1.16)_{1,3}$, we obtain

$$\rho\frac{\partial F}{\partial(y_{j,M})}\frac{d}{dt}(y_{j,M}) + \rho\frac{\partial F}{\partial \mu'_i}\frac{d\mu'_i}{dt} + \rho\frac{\partial F}{\partial(\mu'_{j,M})}\frac{d}{dt}(\mu'_{j,M}) + \rho\frac{\partial F}{\partial E'_i}\frac{dE'_i}{dt} + \rho\frac{\partial F}{\partial \theta}\frac{d\theta}{dt}$$

$$= \tau_{ij}^R X_{M,i}\frac{d}{dt}(y_{j,M}) - {}^R B_j^L \rho\frac{d\mu'_j}{dt} - \rho A_{ij} X_{M,i}\frac{d}{dt}(\mu'_{j,M}) - P_i^R \frac{dE'_i}{dt} - \rho\eta \frac{d\theta}{dt} \tag{5.13}$$

$$+ \lambda\rho\mu'_k \frac{d\mu'_k}{dt} + L_M \rho\left(\mu'_{k,M}\frac{d\mu'_k}{dt} + \mu'_k \frac{d\mu'_{k,M}}{dt}\right),$$

or

$$\left[X_{M,i}\tau_{ij}^R - \rho\frac{\partial F}{\partial(y_{j,M})}\right]\frac{d}{dt}(y_{j,M}) - \left[P_i^R + \rho\frac{\partial F}{\partial E_i'}\right]\frac{dE_i'}{dt} - \rho\left[\eta + \frac{\partial F}{\partial\theta}\right]\frac{d\theta}{dt}$$
$$-\rho\left[{}^R B_i^L - \lambda\mu_i' - L_M\mu_{i,M}' + \frac{\partial F}{\partial\mu_i'}\right]\frac{d\mu_i'}{dt} - \rho\left[X_{M,i}A_{ij} - L_M\mu_j' + \frac{\partial F}{\partial(\mu_{j,M}')}\right]\frac{d}{dt}(\mu_{j,M}') = 0. \quad (5.14)$$

From (5.14), we obtain the reversible parts of the constitutive relations as

$$X_{M,i}A_{ij} = -\frac{\partial F}{\partial(\mu_{j,M}')} + L_M\mu_j', \quad {}^R B_i^L = -\frac{\partial F}{\partial\mu_i'} + \lambda\mu_i' + L_M\mu_{i,M}',$$
$$X_{M,i}\tau_{ij}^R = \rho\frac{\partial F}{\partial(y_{j,M})}, \quad P_i^R = -\rho\frac{\partial F}{\partial E_i'}, \quad \eta = -\frac{\partial F}{\partial\theta}. \quad (5.15)$$

Using (4.2) and (5.15)$_1$, we obtain

$$X_{M,i}A_{ij}\mu_j' = -\frac{\partial F}{\partial(\mu_{j,M}')}\mu_j' + L_M\mu_j'\mu_j' = 0, \quad (5.16)$$

which leads to the following expression of $L_M$:

$$L_M = \frac{1}{\mu_s'^2}\frac{\partial F}{\partial(\mu_{k,M}')}\mu_k'. \quad (5.17)$$

From (1.13)$_2$ we can write

$$\mathbf{B}^L \cdot \mathbf{M}' = {}^R\mathbf{B}^L \cdot \mathbf{M}' + {}^D\mathbf{B}^L \cdot \mathbf{M}' = 0. \quad (5.18)$$

To ensure the satisfaction of (5.18), we require that

$${}^R\mathbf{B}^L \cdot \mathbf{M}' = 0, \quad {}^D\mathbf{B}^L \cdot \mathbf{M}' = 0. \quad (5.19)$$

(5.19)$_1$ and (5.15)$_2$ lead to

$${}^R B_i^L \mu_i' = -\frac{\partial F}{\partial\mu_i'}\mu_i' + \lambda\mu_i'\mu_i' + L_M\mu_i'\mu_{i,M}' = 0. \quad (5.20)$$

From (5.20) we obtain the following expression of $\lambda$:

$$\lambda = \frac{1}{\mu_s'^2}\frac{\partial F}{\partial\mu_k'}\mu_k'. \quad (5.21)$$

The substitution of (5.17) and (5.21) back into (5.15)$_{1,2}$ gives

$$A_{ij} = -y_{i,M}\left[\frac{\partial F}{\partial(\mu_{j,M}')} - \frac{1}{\mu_s'^2}\frac{\partial F}{\partial(\mu_{k,M}')}\mu_k'\mu_j'\right],$$
$${}^R B_i^L = -\frac{\partial F}{\partial\mu_i'} + \frac{1}{\mu_s'^2}\left[\frac{\partial F}{\partial\mu_k'}\mu_k'\mu_i' + \frac{\partial F}{\partial(\mu_{k,M}')}\mu_k'\mu_{i,M}'\right]. \quad (5.22)$$

For the objectivity [19] or rotational invariance of $F$ and (4.6), $F$ can be reduced to a function of the following inner products [6,8]:

$$C_{KL} = y_{i,K}y_{i,L}, \quad G_{LM} = \mu_{i,K}'\mu_{i,M}', \quad N_L = y_{i,L}\mu_i', \quad W_L = y_{i,L}E_i'. \quad (5.23)$$

Instead of the deformation tensor $C_{KL}$, we will use the strain tensor $E_{KL}$. Therefore, we take

$$F = F(E_{KL}; N_K; G_{LM}; W_K; \theta), \quad E_{KL} = \frac{1}{2}(C_{KL} - \delta_{KL}). \quad (5.24)$$

We assume that $F$ is written symmetrically in the following sense:

$$\frac{\partial F}{\partial E_{KL}} = \frac{\partial F}{\partial E_{LK}}, \quad \frac{\partial F}{\partial G_{KL}} = \frac{\partial F}{\partial G_{LK}}. \quad (5.25)$$

In differentiating $F$, the elements of $E_{KL}$ and $G_{KL}$ are treated independently, i.e.,

$$\frac{\partial E_{KL}}{\partial E_{LK}} = 0, \quad \frac{\partial G_{KL}}{\partial G_{LK}} = 0, \quad K \neq L. \tag{5.26}$$

Then, it can be shown that

$$L_M = \frac{\partial F}{\partial (\mu'_{k,M})} \mu'_k = 0. \tag{5.27}$$

The constitutive relations in (5.15)$_{3\text{-}5}$ and (5.22) become

$$\tau^R_{ij} = \rho y_{i,M} \frac{\partial F}{\partial E_{ML}} y_{j,L} + \rho y_{i,M} \frac{\partial F}{\partial N_M} \mu'_j + \rho y_{i,M} \frac{\partial F}{\partial W_M} E'_j, \tag{5.28}$$

$$A_{ij} = -2 y_{i,M} \mu'_{j,L} \frac{\partial F}{\partial G_{ML}}, \tag{5.29}$$

$$^R B^L_i = -y_{i,L} \frac{\partial F}{\partial N_L} + \frac{1}{\mu'^2_s} y_{k,L} \mu'_k \mu'_i \frac{\partial F}{\partial N_L}, \tag{5.30}$$

$$P^R_i = -\rho y_{i,L} \frac{\partial F}{\partial W_L}, \quad \eta = -\frac{\partial F}{\partial \theta}. \tag{5.31}$$

It can be verified that (5.28)-(5.31) satisfy (4.15).

## 6. Recapitulation

In summary, including Maxwell's equations, the field equations are

$$\nabla \cdot \mathbf{D} = \sigma, \tag{6.1}$$

$$\nabla \cdot \mathbf{B} = 0, \tag{6.2}$$

$$\nabla \times \mathbf{E} = -\frac{\partial \mathbf{B}}{\partial t}, \tag{6.3}$$

$$\nabla \times \mathbf{H} = \mathbf{J} + \frac{\partial \mathbf{D}}{\partial t}, \tag{6.4}$$

$$\mathbf{D} = \varepsilon_0 \mathbf{E} + \mathbf{P}, \quad \mathbf{H} = \frac{\mathbf{B}}{\mu_0} - \mathbf{M}, \tag{6.5}$$

$$\frac{1}{\gamma} \frac{d\mu'_i}{dt} = \varepsilon_{ijk} \mu'_j \left( -A_{lk,l} - \frac{A_{lk}}{\rho} \rho_{,l} + B_k + B^L_k \right), \tag{6.6}$$

$$\frac{d\rho}{dt} + \rho \nabla \cdot \mathbf{v} = 0, \tag{6.7}$$

$$\rho \frac{d\mathbf{v}}{dt} = \nabla \cdot \boldsymbol{\tau} + \rho \mathbf{f} + \mathbf{F}^{EM}. \tag{6.8}$$

The constitutive relations are determined by

$$\boldsymbol{\tau} = \boldsymbol{\tau}^R + \boldsymbol{\tau}^D, \quad \mathbf{P} = \mathbf{P}^R + \mathbf{P}^D, \quad \mathbf{B}^L = {}^R\mathbf{B}^L + {}^D\mathbf{B}^L, \tag{6.9}$$

$$F = F(E_{KL}; N_K; G_{LM}; W_K; \theta), \tag{6.10}$$

and expressions for $\mathbf{q}$ and $\mathbf{J}'$. They are restricted by

$${}^D\mathbf{B}^L \cdot \mathbf{M} = 0, \tag{6.11}$$

$$\tau^D_{ij} v_{j,i} - {}^D B^L_k \rho \frac{d\mu'_k}{dt} - P^D_i \frac{dE'_i}{dt} - \frac{q_i}{\theta} \theta_{,i} + J'_i E' \geq 0. \tag{6.12}$$

On the boundary surface of a material body, possible boundary conditions are the prescriptions of [6,8]

$$\begin{aligned}
&\mathbf{y} \quad \text{or} \quad \mathbf{n}\cdot(\boldsymbol{\tau}+\mathbf{T}^{EM}+\mathbf{vG}), \\
&\mathbf{n}\cdot\mathbf{D} \quad \text{and} \quad \mathbf{n}\times\mathbf{H}', \\
&\mathbf{n}\cdot\mathbf{B} \quad \text{and} \quad \mathbf{n}\times\mathbf{E}', \\
&\theta \quad \text{or} \quad \mathbf{n}\cdot\mathbf{q}, \\
&\delta\boldsymbol{\theta} \quad \text{or} \quad \mathbf{n}\cdot\mathbf{A}\times\boldsymbol{\mu}',
\end{aligned} \qquad (6.13)$$

where

$$\mathbf{H}' \cong \mathbf{H} - \mathbf{v}\times\mathbf{D}. \qquad (6.14)$$

## 7. Conclusions

The multi-continuum model with properly chosen charged and magnetized continua is effective in constructing various theories of electrodynamics of deformable continua. The constraint conditions due to magnetic saturation can be taken into constitutive relations effectively by Lagrange multipliers. The above equations represent a generalization of the equations in [6] from insulators to conductors. Similar equations can be found in [20] where they were derived from the charged particle model of continuum electrodynamics. The four-continuum model in the present paper is simpler and reveals the underlying physics from a unique angle.